\documentclass[aps,prb,twocolumn,superscriptaddress]{revtex4-1}

\usepackage{graphicx}

\begin{document}


\title{Unexpected phase locking of magnetic fluctuations in the multi-\textbf{k} magnet USb}


\author{J.  A. Lim}
\email{joshua.ajh.lim@gmail.com}		
\author{E. Blackburn}
\affiliation{School of Physics \& Astronomy, University of Birmingham, Birmingham B15~2TT, UK}

\author{N. Magnani}
\affiliation{Glenn T. Seaborg Center, Chemical Sciences Division, Lawrence Berkeley National Laboratory, Berkeley, CA 94720-8175, USA}

\author{A. Hiess}
\affiliation{European Spallation Source ESS AB - Box 176, 22100 Lund, Sweden}
\affiliation{Institut Laue-Langevin, Bo\^{i}te Postale 156, F-38042 Grenoble, France}

\author{L.-P. Regnault}
\affiliation{SPSMS-MDN, UMR-E CEA/UJF-Grenoble 1, INAC, Grenoble, F-38054, France}

\author{R. Caciuffo}
\author{G. H. Lander}
\affiliation{European Commission, Joint Research Centre, Institute for Transuranium Elements, Postfach 2340, D-76125 Karlsruhe, Germany}

\date{\today}

\begin{abstract}
The spin waves in the multi-\textbf{k} antiferromagnet, USb, soften and become quasielastic well below the AFM ordering temperature, $T_{N}$. This occurs without a magnetic or structural transition. It has been suggested that this change is in fact due to de-phasing of the different multi-\textbf{k} components: a switch from 3-\textbf{k} to 1-\textbf{k} behaviour. In this work, we use inelastic neutron scattering with tri-directional polarisation analysis to probe the quasielastic magnetic excitations and reveal that the 3-\textbf{k} structure does not de-phase. More surprisingly, the paramagnetic correlations also maintain the same clear phase correlations well above T$_{N}$ (up to at least $1.4~T_{N}$). This precursor regime has not been observed before in a multi-\textbf{k} system.
\end{abstract}



\maketitle

\section{Introduction}

The $5f$ electrons in uranium-based compounds often show both localized and itinerant behaviour. At low temperatures, the rocksalt USb looks like a well-localized 3-\textbf{k}  antiferromagnet ($T_{N}=$~213~K) with clearly defined spin waves.

The anisotropy of the spin wave polarization crucially established USb as a multi-\textbf{k} structure, as opposed to a multidomain single-\textbf{k} magnet \cite{Jensen1981}. The 3-\textbf{k} structure is made up of equivalent wavevectors of the form [0~0~1] that add to give spins (which reside on the uranium sites) pointing along local  $\langle$1~1~1$\rangle$-type directions. Within each (0~0~1) plane, the fluctuations of the spins from the equilibrium positions have transverse and longitudinal components with respect to the wavevector of the propagating excitation. For the lower energy mode, the precession of the transverse components is out-of-phase and the corresponding inelastic neutron scattering signal vanishes, whereas the precession of the longitudinal components occurs in-phase and the mode appears to be longitudinally polarized: this forms the acoustic mode. For the higher energy optical mode, the opposite situation occurs and the mode appears as a transverse one.

However, above a temperature $T^{*} \approx$ 150~K ($\sim 0.75~T_{N}$) \emph{inside} the AFM phase, previous studies have shown evidence of changes in the spin wave characteristics. As the material approaches $T^{*}$, the acoustic spin-wave mode becomes heavily damped and collapses in energy \cite{Hagen1988}, with the associated spectral weight remaining finite up to $T_{N}$.  Puzzlingly, there is no structural or magnetic phase transition associated with this mode softening, and cubic symmetry is preserved at all temperatures \cite{Knott1980,Ochiai1994}. $\mu$SR measurements \cite{Asch1994} find a distinct change in the relaxation rate at $T^{*}$, with resistivity data detecting a maximum at this temperature \cite{Schoenes84,Ochiai1994}. 

It has been suggested that this change-over and collapse of the spin waves is due to de-phasing of the individual Fourier components making up the 3-\textbf{k} phase \cite{Asch1994}.

Clear observation of this de-phasing has remained elusive: the $\mu$SR studies are difficult to interpret and lack spatial information \cite{Asch1994}, whilst neutron scattering experiments with uni-directional polarization analysis probe the total projection of the magnetisation onto the plane perpendicular to the scattering vector \textbf{Q} and so are unable to resolve changes in a specific polarization component.

Recently, inelastic neutron scattering experiments with tri-directional polarization analysis have been able to unambiguously confirm the 3-\textbf{k} nature in some materials \cite{Blackburn2005, Magnani2010}. By judicious choice of neutron polarisation directions  relative to the scattering vector and the underlying crystal axes, the scattered intensity is generated by different components of the magnetization fluctuation operator and allows insight into spin-wave excitations in these high symmetry, complex magnetic structures \cite{Caciuffo2007}.

The aim of this work is therefore to use this technique to measure the inelastic spectrum from below $T^{*}$, where there are well defined spin waves, to above $T_{N}$, where there is only quasielastic scattering. By monitoring the change with temperature of the different neutron polarisation channels, and hence polarization of magnetic fluctuations that are particular to the 3-\textbf{k} structure, this gives a clear way to test whether the de-phasing occurs.

\section{Experiment}
USb has an fcc NaCl structure (\textit{a} = 6.197 \AA) and the sample was aligned to access the (1~0~0)-(0~1~0) scattering plane. Only the spin flip (\textit{SF}) channel was measured in order to exclude phonon contributions and detect magnetic scattering at \textbf{Q} = (1, 1, 0) (this magnetic Bragg peak arises from the nuclear zone center at (1~1~1) minus the magnetic  propagation vector (0~0~1)). We define the \textit{xSF} neutron polarisation direction as parallel to \textbf{Q} (i.e. in the [1~1~0] direction) and it is sensitive to the total magnetic fluctuations in the plane perpendicular to \textbf{Q}. \textit{ySF} and \textit{zSF} polarisation channels are along the [$\bar{1}$~1~0] and [0~0~1] crystal directions, respectively. The former probes the component of the magnetization fluctuations along [0~0~1], the latter along [$\bar{1}$~1~0]. 

\begin{figure}  \includegraphics[width=8cm]{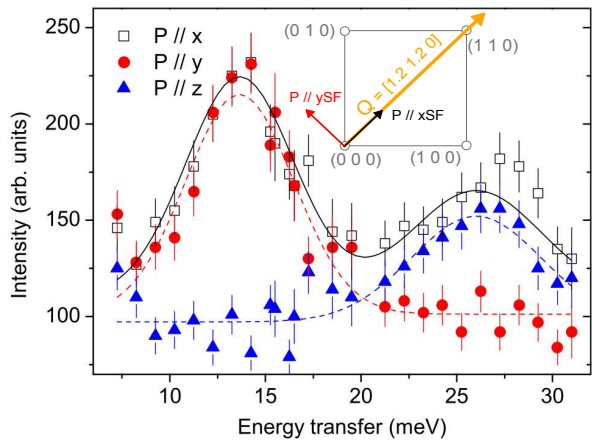}
 \caption{(Color online) Typical inelastic polarised neutron spectra from USb at low temperatures (50~K), along the \textit{x}, \textit{y} and \textit{z} neutron polarization axes (defined in text), adapted from reference \onlinecite{Magnani2010}. The \textit{ySF} and \textit{zSF} channels are able to pick out the acoustic and optical modes, respectively. The total magnetic signal is measured in the \textit{xSF} channel. Inset: a schematic of the sample-polarization geometry.}
 \label{fig:Magnani_result} \end{figure}

As in the previous work by Magnani \textit{et al.} in this geometry \cite{Magnani2010}, the \textit{ySF} neutrons select  the acoustic mode exclusively (as they are sensitive to magnetic fluctuation in the $\langle$0~0~1$\rangle$-type directions), whilst the \textit{zSF} channel selects the optical magnon (sensitive to magnetic fluctuations along [$\bar{1}$~1~0]) (see Fig. \ref{fig:Magnani_result}). Note that although the measurements are at a \textbf{Q} where the Bragg peak is generated by one component, the magnetic fluctuations are sensitive to the total ordered moment, which are along $\langle$1~1~1$\rangle$ in the 3-\textbf{k} state.

Below \textit{T*}, where the 3-\textbf{k} structure has well defined spin waves, the \textit{ySF}/\textit{zSF} intensity ratio will be equal to some maximal value (equal to the flipping ratio, determined by the incident neutron polarisation). However, for spins pointing along the local $\langle$1~1~1$\rangle$-directions in the absence of any phase relationship, the ratio of \textit{ySF}/\textit{zSF} will tend to $\sqrt{2}$ (due to projections of the spins onto the chosen polarisation axes).

This setup can unambiguously resolve the optical and acoustic modes into the different polarization channels and so makes it ideal for testing the de-phasing hypothesis above \textit{T*}, where there are no clearly defined spin waves and the spectral weight becomes quasielastic. Therefore, if the Fourier components de-phase, the relative integrated intensities in the \textit{ySF} and \textit{zSF} channels must also change. 

Polarized neutron measurements were made using a Helmholtz coils setup on  the triple-axis spectrometer IN22 at the Institut Laue-Langevin, Grenoble. The flipping ratio was 13.7 in the \textit{x}, \textit{y} and \textit{z} channels and the energy  resolution was 0.56 meV full width half maximum at the (1,~1,~0) elastic position. The inelastic polarized spectrum was measured at \textbf{Q} = (1,~1,~0) from -2 to 13 meV (we focussed on the acoustic mode only). We measured in fixed final energy mode ($k_{f} = 2.662$ \AA$^{-1}$) and at nine temperatures: from 40 K to 300 K (from below \textit{T*}, to above T$_{N}$).  A single crystal (7~g) of USb was used, the same as in the study by Magnani \textit{et al.} \cite{Magnani2010}.

All inelastic spectra were fitted using the same three components: {\it i}) a two-pole Lorentzian function (for positive and negative energy transfer, details can be found in reference \onlinecite{Schulhof1970}) weighted by the Bose distribution and convoluted with the instrumental resolution function; {\it ii}) an elastic component, centred about the zero of energy, convoluted with the resolution function; {\it iii}) a constant background term.

\section{Results}

\begin{figure}  \includegraphics[width=8cm]{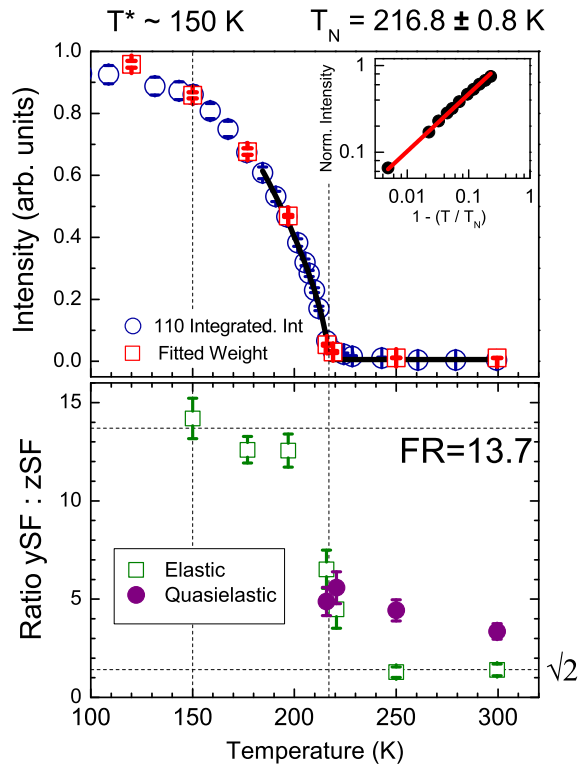}
 \caption{Upper panel: (1, 1, 0) Bragg integrated intensity (blue circles) and the extracted elastic intensities from fitted spectra (red squares) serve as a proxy for the AFM order parameter (fitted with critical exponent - black curve). Inset: log-log plot of the (1,~1,~0) intensity against reduced temperature. Lower panel:  spectra were fitted with elastic and quasielastic components and the ratio of these integrated intensities is plotted vs. temperature. The ratio of the elastic components exactly follows the expected behaviour for the ordered 3-\textbf{k} state, whereas a value of the ratio  between the quasielastic channels greater than $\sqrt{2}$ indicates 3-\textbf{k} correlations.}
 \label{fig:Temp_overview} \end{figure}

A proxy for the AFM order parameter as a function of temperature was obtained by measuring the elastic (1,~1,~0) Bragg intensity and the results are shown in the upper panel of Fig. \ref{fig:Temp_overview}. The effect of extinction is clearly present at the lowest temperatures and no correction for this was made. 

The N\'{e}el temperature ($T_{N} = 216.8 \pm 0.8$~K) and critical exponent ($\beta = 0.33 \pm 0.06$) were extracted, matching published values \cite{Lander1978,Knott1980, Nuttall2002}.

\begin{figure}  \includegraphics[width=8cm]{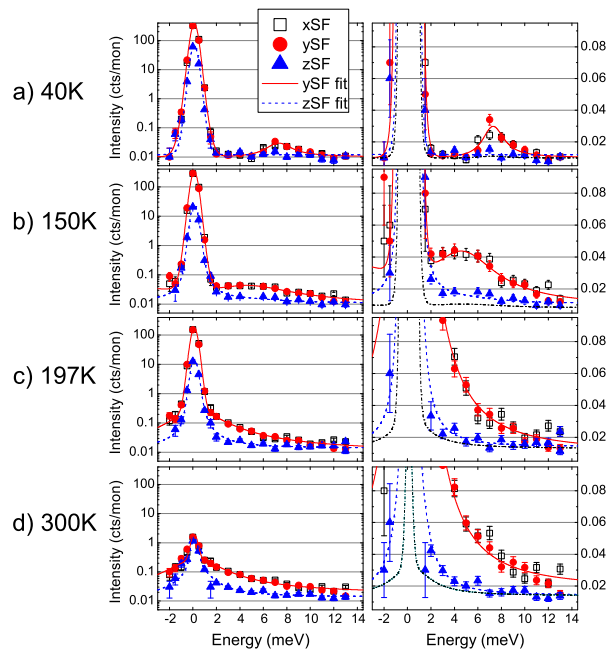}
\caption{Inelastic polarized neutron spectra from the (110) reflection at different temperatures with the experimental data as solid points (with error bars) and lines showing fits to the data. Figures \ref{fig:spectra}(a) \& \ref{fig:spectra}(b) show the spin waves in the \textit{xSF} and \textit{ySF} channels only, which broaden and collapse towards the elastic line with increasing temperature. Figures \ref{fig:spectra}(c) and \ref{fig:spectra}(d) show that the Fourier components do not fully de-phase above $T^{*}$ and  $T_{N}$,  respectively. The black dash-dotted line shows an estimate of the polarisation leakthrough (calculated from the flipping ratio) from the $xSF$ into the $zSF$  channel, showing there is a finite quasielastic component above 40~K.}
\label{fig:spectra} \end{figure}

Inelastic spectra were obtained in the temperature range 40 to 300~K at \textbf{Q}~=~(1,~1,~0). At low temperatures, a well defined spin wave was measured in the \textit{xSF} and \textit{ySF} channels, that broadened and became quasielastic around $T^{*}$ [see Fig. \ref{fig:spectra}(a) \& \ref{fig:spectra}(b)], in excellent quantitative agreement with previous results \cite{Hagen1988, Magnani2010}. We note that in the $zSF$ channel, there is a small amount of  narrow quasielastic broadening at 150~K (and also at 120~K, not shown), which is absent at 40~K. Although fits to this feature and extraction of the integrated intensity were unsuccessful, because of the relatively strong elastic signal; there is a clear quasielastic contribution in the $zSF$ which is present above leakthrough (see dashed-dotted line in Fig. \ref{fig:spectra}). The origin of this scattering is as yet unknown. One possibility might be magnetic domain wall motion appearing in all polarisation channels, which freezes out below T$^{*}$.

Figure \ref{fig:spectra}(c) shows the inelastic polarized neutron spectrum above $T^{*}$ and is characteristic of all spectra in the range $T^{*}$ to $T_{N}$: centred about zero energy transfer is the AFM elastic Bragg peak and a smaller broad quasielastic contribution. 

Here, the intensity in the tails is much greater in the \textit{ySF} channel compared to the \textit{zSF} channel.  This relationship was also observed at 170~K with \textbf{Q}~=~(1.2,~1.2,~0), where magnon-phonon coupling can be safely ignored. The large differences in intensities  between the different polarisation components means that the Fourier components do not fully de-phase above $T^{*}$.

The spectra above $T_{N}$ [e.g. see Fig. \ref{fig:spectra}(d)] no longer shows an elastic contribution from the magnetic Bragg peak and quasielastic broadening is all that remains. Surprisingly, the broad tails show the polarisation conditions remain the same above $T_{N}$, with greater spectral weight in the \textit{ySF} channel relative to \textit{zSF}. This noticeable anisotropy between the \textit{ySF} and \textit{zSF} channels implies phase correlations remain present even in the absence of static magnetic order up to $T = 1.4~T_{N}$.

 \begin{figure}  \includegraphics[width=8cm]{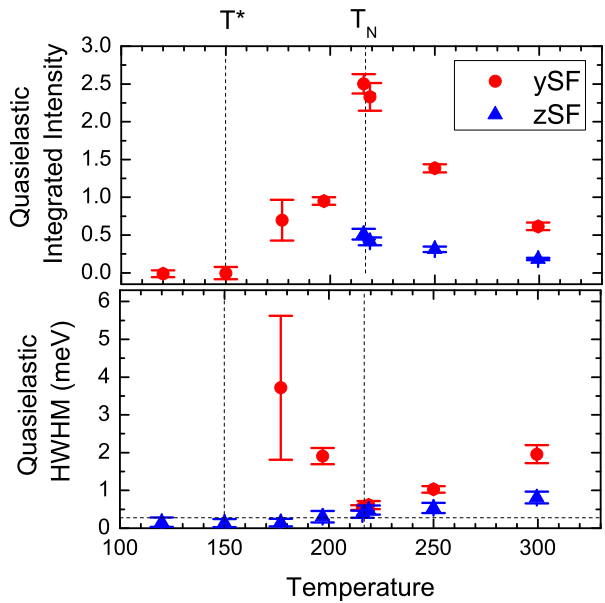}
 \caption{Upper panel: extracted quasielastic integrated intensities. The quasielastic signal develops only above $T^*$ and shows divergent behaviour as expected around $T_N$ in the $ySF$ channel, however this is not seen in the zSF intensity. Lower panel: HWHM of the quasielastic signal, where again critical behaviour is seen in the $ySF$ but not in the $zSF$ (the instrumental resolution is shown by the dashed horizontal line).}
 \label{fig:parameters} \end{figure}

 The quasielastic behaviour can be studied by extracting the fit parameters and the results are shown in Fig. \ref{fig:parameters}. Due to the large elastic signal, the magnitude of the \textit{zSF} quasielastic components could only be reliably extracted at elevated temperatures where the elastic response is diminished.
 
Around $T_{N}$ in the \textit{ySF} channel, the quasielastic integrated intensity diverges and the width becomes narrower. This behaviour is expected and has been previously reported \cite{Hagen1988}. However, the quasielastic features in the \textit{zSF} channel are strikingly different: the feature is much narrower and does not change on passing through the N\'{e}el temperature. This feature will be discussed in more detail later.

For comparison, it is useful to consider the ratio of the \textit{ySF} spectral weight divided by the \textit{zSF} spectral weight. This ratio, split into elastic and quasielastic contributions, is shown as a function of temperature in the lower panel of Fig. \ref{fig:Temp_overview}. 

Below $T_N$, the elastic ratio reaches a maximal value (determined by the incident beam polarization), indicating that the \textit{zSF} elastic intensity is due to leak-through only. Above $T_{N}$,  where there is only a small non-zero elastic component to the fit, the elastic scattering equals the value of $\sqrt{2}$, as expected for incoherent magnetic scattering from spins randomly pointing along local $\langle$1~1~1$\rangle$ directions. The clear agreement of the elastic fit component with theory, indicates the data is well parametrized and the remaining signal is solely quasielastic magnetic scattering.

Above $T^{*}$ in the quasielastic channel, the unequal intensities between the \textit{ySF} and \textit{zSF} channels mean that polarisation conditions are similar to when in the 3-\textbf{k} state (see Fig. \ref{fig:spectra}). Around $T_{N}$ and above, we are able to quantify the relative intensities in the quasielastic ratio  and show this phase relationship remains and, whilst decreasing slowly, appears robust to $\sim$100~K above $T_{N}$ (see lower panel of Fig. \ref{fig:Temp_overview}). The quasielastic signal cannot be from phonon leakthrough as this spectral weight should increase with temperature.

This result also confirms that the correlations above the N\'{e}el temperature, seen previously in neutron scattering and bulk measurements,  are indeed magnetic in origin \cite{Hagen1988, Ochiai1994}.

\section{Discussion}

This  discussion contains two parts: the observed behaviour between  $T^{*}$ and $T_{N}$, where the spin wave becomes quasielastic with sample still AFM ordered, and the behaviour above $T_{N}$, where there are correlations in the neutron polarisation data in the absence of long range-order.

The collapse of a spin wave mode to zero energy is often an indicator that a system is unstable against fluctuations and an associated magnetic phase transition is nearby - a feature seen in many materials. It is therefore unusual that on approaching $T^{*}$, no transition is seen - either in the specific heat data or in the nuclear or magnetic structure \cite{Ochiai1994,Knott1980,Hagen1988}. A possible explanation for this change is that the spins, making up the 3-\textbf{k} structure,  de-phase with one another. We have been able to test this hypothesis using polarized neutrons and conclude that it does not occur.

We attribute a non-$\sqrt{2}$ value of the quasielastic ratio (i.e. greater than expected intensity in the $ySF$ over the  \textit{zSF} channel) as evidence for 3-\textbf{k} correlations. This interpretation is in good agreement with theoretical and experimental results at low temperatures \cite{Jensen1981,Magnani2010} and can be generalized to higher temperatures. Hence, if the spins are  uncorrelated or de-phase with one another, then the ratio should tend to $\sqrt{2}$. Conversely if some correlation persists, the ratio should be greater than $\sqrt{2}$.

Whilst we are unable to explain the cause of the mode softening at $T^{*}$, we suggest understanding of the behaviour may rely on the itinerant-local duality found in many $5f$ electron systems. In particular we point towards  a more itinerant picture of USb, considering the quasielastic nature of the spin waves. Whilst much work has been done on measuring and modelling the low  temperature behaviour (e.g. dHvA, SDFT, etc. \cite{Hotta1995, Knopfle2000, Yamagami2000}), more study is needed to understand the high temperature behaviour \cite{Ochiai1994,Kumigashira2000}.

We note that polarized inelastic neutron scattering is unable to discern whether any partial de-phasing is due to a mixture of fully phase-locked and fully de-phased 3-\textbf{k} structure or partial de-phasing of the whole magnet; however, one would not expect a coexistence of locked and unlocked spins considering the large temperature range over which this behaviour is seen ($\sim 60$~K).

At low temperatures, the physical properties of USb are very well described by a mean-field model where the electrons are localized and the exchange interaction dominates, thus generating large magnetic moments pointing along the cubic cell diagonal. On the other hand, the ARPES band structure seems more consistent with an interpretation based on itinerant electrons \cite{Kumigashira2000}. The possible transition toward a larger degree of itinerancy at high temperatures might also help to explain some peculiar bulk properties, such as the resistivity peak \cite{Schoenes84,Ochiai1994} corresponding to $T^{*}$. The spin wave becoming quasielastic at $T^{*}$ is clearly an indication that the transition between the localized exchange levels, which gives rise to the observed excitations, can no longer be understood from a mean-field point of view at such high temperatures. If so, and considering that the present work disproves both the dephasing hypothesis and any involvement of interactions with phonons, it seems natural to propose that the softening is linked to a higher degree of itinerancy, which could reduce the value of the magnetic moment, broaden the spin-wave transition and therefore also lower its energy range.

The measurements above the N\'{e}el temperature are also striking and unexpected. As the material exits the ordered AFM phase, there is little change in the \textit{zSF} quasielastic scattering and the $ySF$:$zSF$ ratio remains distinctly different from $\sqrt{2}$ (see lower panel of Fig. \ref{fig:Temp_overview}). This suggests that even in the absence of long-range order, the spins maintain a strong, partial phase relationship with one another. Indeed, the presence of these phase correlations may indeed be important in forming the multi-\textbf{k} state in USb.

It should be noted that this precursor regime (above $T_{N}$) shows persistent 3-\textbf{k}-like correlations, which have not been previously reported. It is surprising that these correlations should extend to such high temperatures; particularly as it is the fourth order terms in a Landau free energy expansion that are needed to stabilize the 3-\textbf{k} over the single-\textbf{k} state. 

It could be interesting to realize whether the strong exchange interaction, which dominates at low temperatures, is itself pushing the system toward stronger localization or if the magnetic properties are simply reflecting the effect of another driving force. The peculiar properties of the 3-{\bf k} structure allow us to observe directly the magnetic response of the itinerant states in the $zSF$ channel which, remarkably, does not show any critical behaviour at $T_{N}$. At a glance, the interplay between these itinerant states and the localized moment seems somewhat indirect, with the appearance of the former linked to $T^{*}$ and the disappearance of the latter linked to $T_{N}$; on the other hand, the effect of the exchange interaction on the high-temperature behaviour is still clearly visible in the strong phase correlations.

\section{Conclusions}
The inelastic spectra of the multi-\textbf{k} antiferromagnet USb have been explored using polarised neutrons, across a broad range of temperatures. At the lowest temperatures, we observe a well defined localised 3-\textbf{k} magnet. However, tri-directional polarisation analysis has shown that, contrary to prediction, the Fourier components do not de-phase above $T^{*}$. The cause of the spin wave softening at  $T^{*}$ still remains an open question. 

Surprisingly, USb maintains phase correlations to at least $1.4~T_{N}$ in its quasielastic excitations, despite possessing no long range magnetic order. This precursor region above the N\'{e}el temperature is unexpected but may be important for the formation of the 3-\textbf{k} state in USb.

\begin{acknowledgments}
The authors would like to thank E. M. Forgan for comments on the manuscript and  Dr. O. Vogt. of ETH, Zurich  for providing the sample.

\end{acknowledgments}

\bibliography{Multi-k_refs,USb}

\begin{thebibliography}{16}%
\makeatletter
\providecommand \@ifxundefined [1]{%
 \@ifx{#1\undefined}
}%
\providecommand \@ifnum [1]{%
 \ifnum #1\expandafter \@firstoftwo
 \else \expandafter \@secondoftwo
 \fi
}%
\providecommand \@ifx [1]{%
 \ifx #1\expandafter \@firstoftwo
 \else \expandafter \@secondoftwo
 \fi
}%
\providecommand \natexlab [1]{#1}%
\providecommand \enquote  [1]{``#1''}%
\providecommand \bibnamefont  [1]{#1}%
\providecommand \bibfnamefont [1]{#1}%
\providecommand \citenamefont [1]{#1}%
\providecommand \href@noop [0]{\@secondoftwo}%
\providecommand \href [0]{\begingroup \@sanitize@url \@href}%
\providecommand \@href[1]{\@@startlink{#1}\@@href}%
\providecommand \@@href[1]{\endgroup#1\@@endlink}%
\providecommand \@sanitize@url [0]{\catcode `\\12\catcode `\$12\catcode
  `\&12\catcode `\#12\catcode `\^12\catcode `\_12\catcode `\%12\relax}%
\providecommand \@@startlink[1]{}%
\providecommand \@@endlink[0]{}%
\providecommand \url  [0]{\begingroup\@sanitize@url \@url }%
\providecommand \@url [1]{\endgroup\@href {#1}{\urlprefix }}%
\providecommand \urlprefix  [0]{URL }%
\providecommand \Eprint [0]{\href }%
\providecommand \doibase [0]{http://dx.doi.org/}%
\providecommand \selectlanguage [0]{\@gobble}%
\providecommand \bibinfo  [0]{\@secondoftwo}%
\providecommand \bibfield  [0]{\@secondoftwo}%
\providecommand \translation [1]{[#1]}%
\providecommand \BibitemOpen [0]{}%
\providecommand \bibitemStop [0]{}%
\providecommand \bibitemNoStop [0]{.\EOS\space}%
\providecommand \EOS [0]{\spacefactor3000\relax}%
\providecommand \BibitemShut  [1]{\csname bibitem#1\endcsname}%
\let\auto@bib@innerbib\@empty
\bibitem [{\citenamefont {Jensen}\ and\ \citenamefont
  {Bak}(1981)}]{Jensen1981}%
  \BibitemOpen
  \bibfield  {author} {\bibinfo {author} {\bibfnamefont {J.}~\bibnamefont
  {Jensen}}\ and\ \bibinfo {author} {\bibfnamefont {P.}~\bibnamefont {Bak}},\
  }\href {\doibase 10.1103/PhysRevB.23.6180} {\bibfield  {journal} {\bibinfo
  {journal} {Phys. Rev. B}\ }\textbf {\bibinfo {volume} {23}},\ \bibinfo
  {pages} {6180} (\bibinfo {year} {1981})}\BibitemShut {NoStop}%
\bibitem [{\citenamefont {Hagen}\ \emph {et~al.}(1988)\citenamefont {Hagen},
  \citenamefont {Stirling},\ and\ \citenamefont {Lander}}]{Hagen1988}%
  \BibitemOpen
  \bibfield  {author} {\bibinfo {author} {\bibfnamefont {M.}~\bibnamefont
  {Hagen}}, \bibinfo {author} {\bibfnamefont {W.~G.}\ \bibnamefont {Stirling}},
  \ and\ \bibinfo {author} {\bibfnamefont {G.~H.}\ \bibnamefont {Lander}},\
  }\href {\doibase 10.1103/PhysRevB.37.1846} {\bibfield  {journal} {\bibinfo
  {journal} {Phys. Rev. B}\ }\textbf {\bibinfo {volume} {37}},\ \bibinfo
  {pages} {1846} (\bibinfo {year} {1988})}\BibitemShut {NoStop}%
\bibitem [{\citenamefont {Knott}\ \emph {et~al.}(1980)\citenamefont {Knott},
  \citenamefont {Lander}, \citenamefont {Mueller},\ and\ \citenamefont
  {Vogt}}]{Knott1980}%
  \BibitemOpen
  \bibfield  {author} {\bibinfo {author} {\bibfnamefont {H.~W.}\ \bibnamefont
  {Knott}}, \bibinfo {author} {\bibfnamefont {G.~H.}\ \bibnamefont {Lander}},
  \bibinfo {author} {\bibfnamefont {M.~H.}\ \bibnamefont {Mueller}}, \ and\
  \bibinfo {author} {\bibfnamefont {O.}~\bibnamefont {Vogt}},\ }\href {\doibase
  10.1103/PhysRevB.21.4159} {\bibfield  {journal} {\bibinfo  {journal} {Phys.
  Rev. B}\ }\textbf {\bibinfo {volume} {21}},\ \bibinfo {pages} {4159}
  (\bibinfo {year} {1980})}\BibitemShut {NoStop}%
\bibitem [{\citenamefont {Ochiai}\ \emph {et~al.}(1994)\citenamefont {Ochiai},
  \citenamefont {Suzuki}, \citenamefont {Shikama}, \citenamefont {Suzuki},
  \citenamefont {Hotta}, \citenamefont {Haga},\ and\ \citenamefont
  {Suzuki}}]{Ochiai1994}%
  \BibitemOpen
  \bibfield  {author} {\bibinfo {author} {\bibfnamefont {A.}~\bibnamefont
  {Ochiai}}, \bibinfo {author} {\bibfnamefont {Y.}~\bibnamefont {Suzuki}},
  \bibinfo {author} {\bibfnamefont {T.}~\bibnamefont {Shikama}}, \bibinfo
  {author} {\bibfnamefont {K.}~\bibnamefont {Suzuki}}, \bibinfo {author}
  {\bibfnamefont {E.}~\bibnamefont {Hotta}}, \bibinfo {author} {\bibfnamefont
  {Y.}~\bibnamefont {Haga}}, \ and\ \bibinfo {author} {\bibfnamefont
  {T.}~\bibnamefont {Suzuki}},\ }\href {\doibase 10.1016/0921-4526(94)91923-2}
  {\bibfield  {journal} {\bibinfo  {journal} {Physica B: Condensed Matter}\
  }\textbf {\bibinfo {volume} {199}},\ \bibinfo {pages} {616 } (\bibinfo {year}
  {1994})}\BibitemShut {NoStop}%
\bibitem [{\citenamefont {Asch}\ \emph {et~al.}(1994)\citenamefont {Asch},
  \citenamefont {Kalvius}, \citenamefont {Kratzer},\ and\ \citenamefont
  {Litterst}}]{Asch1994}%
  \BibitemOpen
  \bibfield  {author} {\bibinfo {author} {\bibfnamefont {L.}~\bibnamefont
  {Asch}}, \bibinfo {author} {\bibfnamefont {G.~M.}\ \bibnamefont {Kalvius}},
  \bibinfo {author} {\bibfnamefont {A.}~\bibnamefont {Kratzer}}, \ and\
  \bibinfo {author} {\bibfnamefont {F.~J.}\ \bibnamefont {Litterst}},\ }\href
  {http://dx.doi.org/10.1007/BF02069420} {\bibfield  {journal} {\bibinfo
  {journal} {Hyperfine Interactions}\ }\textbf {\bibinfo {volume} {85}},\
  \bibinfo {pages} {193} (\bibinfo {year} {1994})}\BibitemShut {NoStop}%
\bibitem [{\citenamefont {Schoenes}\ \emph {et~al.}(1984)\citenamefont
  {Schoenes}, \citenamefont {Frick},\ and\ \citenamefont {Vogt}}]{Schoenes84}%
  \BibitemOpen
  \bibfield  {author} {\bibinfo {author} {\bibfnamefont {J.}~\bibnamefont
  {Schoenes}}, \bibinfo {author} {\bibfnamefont {B.}~\bibnamefont {Frick}}, \
  and\ \bibinfo {author} {\bibfnamefont {O.}~\bibnamefont {Vogt}},\ }\href
  {\doibase 10.1103/PhysRevB.30.6578} {\bibfield  {journal} {\bibinfo
  {journal} {Phys. Rev. B}\ }\textbf {\bibinfo {volume} {30}},\ \bibinfo
  {pages} {6578} (\bibinfo {year} {1984})}\BibitemShut {NoStop}%
\bibitem [{\citenamefont {Blackburn}\ \emph {et~al.}(2005)\citenamefont
  {Blackburn}, \citenamefont {Caciuffo}, \citenamefont {Magnani}, \citenamefont
  {Santini}, \citenamefont {Brown}, \citenamefont {Enderle},\ and\
  \citenamefont {Lander}}]{Blackburn2005}%
  \BibitemOpen
  \bibfield  {author} {\bibinfo {author} {\bibfnamefont {E.}~\bibnamefont
  {Blackburn}}, \bibinfo {author} {\bibfnamefont {R.}~\bibnamefont {Caciuffo}},
  \bibinfo {author} {\bibfnamefont {N.}~\bibnamefont {Magnani}}, \bibinfo
  {author} {\bibfnamefont {P.}~\bibnamefont {Santini}}, \bibinfo {author}
  {\bibfnamefont {P.~J.}\ \bibnamefont {Brown}}, \bibinfo {author}
  {\bibfnamefont {M.}~\bibnamefont {Enderle}}, \ and\ \bibinfo {author}
  {\bibfnamefont {G.~H.}\ \bibnamefont {Lander}},\ }\href {\doibase
  10.1103/PhysRevB.72.184411} {\bibfield  {journal} {\bibinfo  {journal} {Phys.
  Rev. B}\ }\textbf {\bibinfo {volume} {72}},\ \bibinfo {pages} {184411}
  (\bibinfo {year} {2005})}\BibitemShut {NoStop}%
\bibitem [{\citenamefont {Magnani}\ \emph {et~al.}(2010)\citenamefont
  {Magnani}, \citenamefont {Caciuffo}, \citenamefont {Lander}, \citenamefont
  {Hiess},\ and\ \citenamefont {Regnault}}]{Magnani2010}%
  \BibitemOpen
  \bibfield  {author} {\bibinfo {author} {\bibfnamefont {N.}~\bibnamefont
  {Magnani}}, \bibinfo {author} {\bibfnamefont {R.}~\bibnamefont {Caciuffo}},
  \bibinfo {author} {\bibfnamefont {G.~H.}\ \bibnamefont {Lander}}, \bibinfo
  {author} {\bibfnamefont {A.}~\bibnamefont {Hiess}}, \ and\ \bibinfo {author}
  {\bibfnamefont {L.-P.}\ \bibnamefont {Regnault}},\ }\href
  {http://stacks.iop.org/0953-8984/22/i=11/a=116002} {\bibfield  {journal}
  {\bibinfo  {journal} {Journal of Physics: Condensed Matter}\ }\textbf
  {\bibinfo {volume} {22}},\ \bibinfo {pages} {116002} (\bibinfo {year}
  {2010})}\BibitemShut {NoStop}%
\bibitem [{\citenamefont {Caciuffo}\ \emph {et~al.}(2007)\citenamefont
  {Caciuffo}, \citenamefont {Magnani}, \citenamefont {Santini}, \citenamefont
  {Carretta}, \citenamefont {Amoretti}, \citenamefont {Blackburn},
  \citenamefont {Enderle}, \citenamefont {Brown},\ and\ \citenamefont
  {Lander}}]{Caciuffo2007}%
  \BibitemOpen
  \bibfield  {author} {\bibinfo {author} {\bibfnamefont {R.}~\bibnamefont
  {Caciuffo}}, \bibinfo {author} {\bibfnamefont {N.}~\bibnamefont {Magnani}},
  \bibinfo {author} {\bibfnamefont {P.}~\bibnamefont {Santini}}, \bibinfo
  {author} {\bibfnamefont {S.}~\bibnamefont {Carretta}}, \bibinfo {author}
  {\bibfnamefont {G.}~\bibnamefont {Amoretti}}, \bibinfo {author}
  {\bibfnamefont {E.}~\bibnamefont {Blackburn}}, \bibinfo {author}
  {\bibfnamefont {M.}~\bibnamefont {Enderle}}, \bibinfo {author} {\bibfnamefont
  {P.}~\bibnamefont {Brown}}, \ and\ \bibinfo {author} {\bibfnamefont
  {G.}~\bibnamefont {Lander}},\ }\href {\doibase 10.1016/j.jmmm.2006.10.536}
  {\bibfield  {journal} {\bibinfo  {journal} {Journal of Magnetism and Magnetic
  Materials}\ }\textbf {\bibinfo {volume} {310}},\ \bibinfo {pages} {1698 }
  (\bibinfo {year} {2007})}\BibitemShut {NoStop}%
\bibitem [{\citenamefont {Schulhof}\ \emph {et~al.}(1970)\citenamefont
  {Schulhof}, \citenamefont {Heller}, \citenamefont {Nathans},\ and\
  \citenamefont {Linz}}]{Schulhof1970}%
  \BibitemOpen
  \bibfield  {author} {\bibinfo {author} {\bibfnamefont {M.~P.}\ \bibnamefont
  {Schulhof}}, \bibinfo {author} {\bibfnamefont {P.}~\bibnamefont {Heller}},
  \bibinfo {author} {\bibfnamefont {R.}~\bibnamefont {Nathans}}, \ and\
  \bibinfo {author} {\bibfnamefont {A.}~\bibnamefont {Linz}},\ }\href {\doibase
  10.1103/PhysRevB.1.2304} {\bibfield  {journal} {\bibinfo  {journal} {Phys.
  Rev. B}\ }\textbf {\bibinfo {volume} {1}},\ \bibinfo {pages} {2304} (\bibinfo
  {year} {1970})}\BibitemShut {NoStop}%
\bibitem [{\citenamefont {Lander}\ \emph {et~al.}(1978)\citenamefont {Lander},
  \citenamefont {Sinha}, \citenamefont {Sparlin},\ and\ \citenamefont
  {Vogt}}]{Lander1978}%
  \BibitemOpen
  \bibfield  {author} {\bibinfo {author} {\bibfnamefont {G.~H.}\ \bibnamefont
  {Lander}}, \bibinfo {author} {\bibfnamefont {S.~K.}\ \bibnamefont {Sinha}},
  \bibinfo {author} {\bibfnamefont {D.~M.}\ \bibnamefont {Sparlin}}, \ and\
  \bibinfo {author} {\bibfnamefont {O.}~\bibnamefont {Vogt}},\ }\href {\doibase
  10.1103/PhysRevLett.40.523} {\bibfield  {journal} {\bibinfo  {journal} {Phys.
  Rev. Lett.}\ }\textbf {\bibinfo {volume} {40}},\ \bibinfo {pages} {523}
  (\bibinfo {year} {1978})}\BibitemShut {NoStop}%
\bibitem [{\citenamefont {Nuttall}\ \emph {et~al.}(2002)\citenamefont
  {Nuttall}, \citenamefont {Perry}, \citenamefont {Stirling}, \citenamefont
  {Mitchell}, \citenamefont {Kilcoyne},\ and\ \citenamefont
  {Cywinski}}]{Nuttall2002}%
  \BibitemOpen
  \bibfield  {author} {\bibinfo {author} {\bibfnamefont {W.}~\bibnamefont
  {Nuttall}}, \bibinfo {author} {\bibfnamefont {S.}~\bibnamefont {Perry}},
  \bibinfo {author} {\bibfnamefont {W.}~\bibnamefont {Stirling}}, \bibinfo
  {author} {\bibfnamefont {P.}~\bibnamefont {Mitchell}}, \bibinfo {author}
  {\bibfnamefont {S.}~\bibnamefont {Kilcoyne}}, \ and\ \bibinfo {author}
  {\bibfnamefont {R.}~\bibnamefont {Cywinski}},\ }\href {\doibase
  10.1016/S0921-4526(01)01034-1} {\bibfield  {journal} {\bibinfo  {journal}
  {Physica B: Condensed Matter}\ }\textbf {\bibinfo {volume} {315}},\ \bibinfo
  {pages} {179 } (\bibinfo {year} {2002})}\BibitemShut {NoStop}%
\bibitem [{\citenamefont {Hotta}\ \emph {et~al.}(1995)\citenamefont {Hotta},
  \citenamefont {Ochiai}, \citenamefont {Suzuki}, \citenamefont {Shikama},
  \citenamefont {Suzuki}, \citenamefont {Haga},\ and\ \citenamefont
  {Suzuki}}]{Hotta1995}%
  \BibitemOpen
  \bibfield  {author} {\bibinfo {author} {\bibfnamefont {E.}~\bibnamefont
  {Hotta}}, \bibinfo {author} {\bibfnamefont {A.}~\bibnamefont {Ochiai}},
  \bibinfo {author} {\bibfnamefont {Y.}~\bibnamefont {Suzuki}}, \bibinfo
  {author} {\bibfnamefont {T.}~\bibnamefont {Shikama}}, \bibinfo {author}
  {\bibfnamefont {K.}~\bibnamefont {Suzuki}}, \bibinfo {author} {\bibfnamefont
  {Y.}~\bibnamefont {Haga}}, \ and\ \bibinfo {author} {\bibfnamefont
  {T.}~\bibnamefont {Suzuki}},\ }\href {\doibase 10.1016/0925-8388(94)05002-3}
  {\bibfield  {journal} {\bibinfo  {journal} {Journal of Alloys and Compounds}\
  }\textbf {\bibinfo {volume} {219}},\ \bibinfo {pages} {252 } (\bibinfo {year}
  {1995})}\BibitemShut {NoStop}%
\bibitem [{\citenamefont {Kn\"opfle}\ and\ \citenamefont
  {Sandratskii}(2000)}]{Knopfle2000}%
  \BibitemOpen
  \bibfield  {author} {\bibinfo {author} {\bibfnamefont {K.}~\bibnamefont
  {Kn\"opfle}}\ and\ \bibinfo {author} {\bibfnamefont {L.~M.}\ \bibnamefont
  {Sandratskii}},\ }\href {\doibase 10.1103/PhysRevB.63.014411} {\bibfield
  {journal} {\bibinfo  {journal} {Phys. Rev. B}\ }\textbf {\bibinfo {volume}
  {63}},\ \bibinfo {pages} {014411} (\bibinfo {year} {2000})}\BibitemShut
  {NoStop}%
\bibitem [{\citenamefont {Yamagami}(2000)}]{Yamagami2000}%
  \BibitemOpen
  \bibfield  {author} {\bibinfo {author} {\bibfnamefont {H.}~\bibnamefont
  {Yamagami}},\ }\href {\doibase 10.1103/PhysRevB.61.6246} {\bibfield
  {journal} {\bibinfo  {journal} {Phys. Rev. B}\ }\textbf {\bibinfo {volume}
  {61}},\ \bibinfo {pages} {6246} (\bibinfo {year} {2000})}\BibitemShut
  {NoStop}%
\bibitem [{\citenamefont {Kumigashira}\ \emph {et~al.}(2000)\citenamefont
  {Kumigashira}, \citenamefont {Ito}, \citenamefont {Ashihara}, \citenamefont
  {Kim}, \citenamefont {Aoki}, \citenamefont {Suzuki}, \citenamefont
  {Yamagami}, \citenamefont {Takahashi},\ and\ \citenamefont
  {Ochiai}}]{Kumigashira2000}%
  \BibitemOpen
  \bibfield  {author} {\bibinfo {author} {\bibfnamefont {H.}~\bibnamefont
  {Kumigashira}}, \bibinfo {author} {\bibfnamefont {T.}~\bibnamefont {Ito}},
  \bibinfo {author} {\bibfnamefont {A.}~\bibnamefont {Ashihara}}, \bibinfo
  {author} {\bibfnamefont {H.-D.}\ \bibnamefont {Kim}}, \bibinfo {author}
  {\bibfnamefont {H.}~\bibnamefont {Aoki}}, \bibinfo {author} {\bibfnamefont
  {T.}~\bibnamefont {Suzuki}}, \bibinfo {author} {\bibfnamefont
  {H.}~\bibnamefont {Yamagami}}, \bibinfo {author} {\bibfnamefont
  {T.}~\bibnamefont {Takahashi}}, \ and\ \bibinfo {author} {\bibfnamefont
  {A.}~\bibnamefont {Ochiai}},\ }\href {\doibase 10.1103/PhysRevB.61.15707}
  {\bibfield  {journal} {\bibinfo  {journal} {Phys. Rev. B}\ }\textbf {\bibinfo
  {volume} {61}},\ \bibinfo {pages} {15707} (\bibinfo {year}
  {2000})}\BibitemShut {NoStop}%
\end{thebibliography}%

\end{document}